\documentclass{aa501}

\usepackage{graphicx}
\usepackage{natbib}

\newcommand{\id}{\mbox{$\mathrm{d^{-1}}$}}
\newcommand{\Msun}{\mbox{$M_{\odot}$}}
\newcommand{\kms}{\mbox{$\mathrm{km\,s^{-1}}$}}
\newcommand{\Mwd}{\mbox{$M_\mathrm{wd}$}}
\newcommand{\Msec}{\mbox{$M_\mathrm{sec}$}}
\newcommand{\Rsec}{\mbox{$R_\mathrm{sec}$}}
\newcommand{\Ksec}{\mbox{$K_\mathrm{sec}$}}
\newcommand{\Teff}{\mbox{$T_\mathrm{eff}$}}
\newcommand{\forb}{\mbox{$f_{\rm orb}$}}
\newcommand{\Porb}{\mbox{$P_{\rm orb}$}}

\newcommand{\Ha}{\mbox{${\mathrm H\alpha}$}}
\newcommand{\Hb}{\mbox{${\mathrm H\beta}$}}

\newcommand{\Hd}{\mbox{${\mathrm H\delta}$}}

\begin{document}

\title{HS\,2237+8154: On the onset of mass transfer \\or entering the
period gap?\thanks{Based in part on observations made at
the 1.2m telescope, located at Kryoneri Korinthias, and owned by the
National Observatory of Athens, Greece, and with the Isaac Newton
Telescope, which is operated on the island of La Palma by the Isaac
Newton Group in the Spanish Observatorio del Roque de los Muchachos of
the Instituto de Astrofisica de Canarias.}}

\author{B.T. G\"ansicke\inst{1,2} \and
        S. Araujo-Betancor\inst{1}\and
        H.-J. Hagen\inst{3} \and
        E. Harlaftis\inst{4} \and
        S. Kitsionas\inst{5} \and
        S. Dreizler\inst{6} \thanks{Visiting Astronomer,
         German-Spanish Astronomical Centre, Calar Alto, operated by
         the Max-Planck-Institute for Astronomy, Heidelberg, jointly
         with the Spanish National Commission for Astronomy.} \and
        D. Engels\inst{3}
        }   
\offprints{B.T. G\"ansicke, e-mail: Boris.Gaensicke@warwick.ac.uk}

\institute{
  School of Physics and Astronomy, University of Southampton,
  Southampton SO17 1BJ, UK
\and
  Department of Physics, University of Warwick, Coventry CV4 7AL, UK
\and
  Hamburger Sternwarte, Universit\"at Hamburg, Gojenbergsweg 112,
  21029 Hamburg, Germany
\and
  Institute of Space Applications and Remote Sensing, National
  Observatory of Athens, P.O. Box 20048, Athens 11810, Greece
\and
  Institute of Astronomy and Astrophysics, National Observatory of Athens,
  P.O. Box 20048, Athens 11810, Greece
\and
Universit\"ats-Sternwarte, Geismarlandstr. 11, D-37083 G\"ottingen, Germany
}

\date{Received \underline{\hskip2cm} ; accepted 15 January 2003 }

\abstract{We report follow-up observations of a new white dwarf/red
dwarf binary HS\,2237+8154, identified as a blue variable star from
the Hamburg Quasar Survey.  Ellipsoidal modulation observed in the $R$
band as well as the $\Ha$ radial velocity variations measured from
time-resolved spectroscopy determine the orbital period to be
$\Porb=178.10\pm0.08$\,min. The optical spectrum of HS\,2237+8154 is
well described by a combination of a $\Teff=11\,500\pm1500$\,K white
dwarf (assuming $\log g=8$) and a dM3.5$\pm$0.5 secondary star. The
distance implied from the flux scaling factors of both stellar
components is $d=105\pm25$\,pc. Combining the constraints obtained from
the radial velocity of the secondary and from the ellipsoidal modulation,
we derive a binary inclination of $i\simeq50^{\circ}-70^{\circ}$ and
stellar masses of $\Mwd=0.47-0.67$\,\Msun\ and
$\Msec=0.2-0.4$\,\Msun.  All observations imply that the secondary
star must be nearly Roche-lobe filling. Consequently, HS\,2237+8154
may be either a pre-cataclysmic variable close to the start of mass
transfer, or~--~considering its orbital period~--~a cataclysmic
variable that terminated mass transfer and entered the period gap, or
a hibernating nova.
 } 

\maketitle

\section{Introduction}
The currently known population of cataclysmic variables (CVs)
comprises more than 1000 systems, with period measurements available
for $\simeq500$ of them \citep{downesetal01-1, kubeetal03-1,
ritter+kolb03-1}. Compared to these numbers, only very few CV
progenitors have been identified so
far. \citet{schreiber+gaensicke03-1} have recently analysed 30
well-observed pre-CVs and found that the sample of known pre-CVs is
strongly biased towards systems containing hot (young) white dwarfs
and late type (low-mass) companion stars. Testing and improving our
understanding of the evolution of pre-CVs through the common envelope
phase and of the angular momentum loss mechisms that subsequently
brings them into a semidetached configuration will enormously benefit
from a larger and unbiased sample of such stars.

We have identified a detached white dwarf/red dwarf binary
(Fig.\,\ref{f-fc}) in the course of our studies of the stellar content
of the Hamburg Quasar Survey \citep[e.g.][]{gaensickeetal00-2,
nogamietal00-1, szkodyetal01-1, gaensickeetal02-2,
araujo-betancoretal03-2}. Our follow-up observations show that
HS\,2237+8154 may belong to the large population of ``cold \& old''
pre-CVs predicted by \citet{schreiber+gaensicke03-1}, however, it may
as well be a CV in a peculiar phase of its evolution.

\section{Observations and Data Reduction}

\subsection{Spectroscopy}
\begin{figure}
\includegraphics[angle=-90,width=8.8cm]{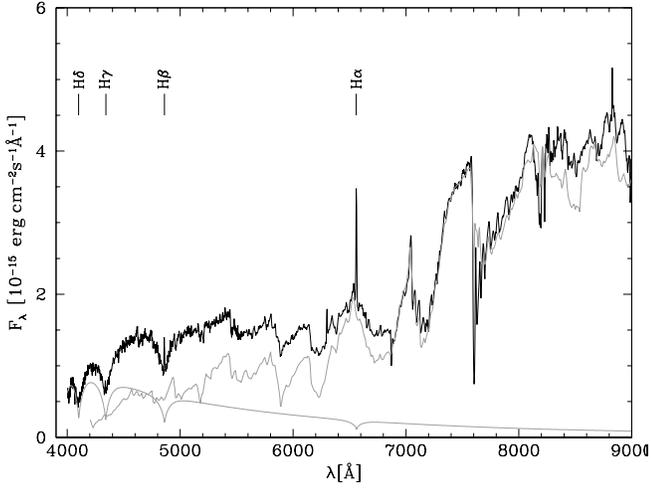}
\caption[]{\label{f-idspec} The Calar Alto TWIN idenfication spectrum
of HS\,2237+8154. Plotted in gray is a two-component fit consisting of a
white dwarf model ($\Teff=12\,000$\,K, $\log g=8.0$) and a dM3.5
template (Gl273). }
\end{figure}

A single identification spectrum of HS\,2237+8154 was obtained in June
1991 using the 3.5m telescope at Calar Alto equipped with the TWIN
spectrograph (Table\,\ref{t-obslog}). Gratings with 144 and
160\,\AA/mm were used in the blue and red channel respectively
providing a resolution of about 6\,\AA. Standard data reduction with
flat fielding, bias subtraction as well as cosmic-ray filtering was
performed in Kiel using IDAS, a long-slit spectroscopy package written
by G. Jonas (University Kiel). The TWIN spectrum
(Fig.\,\ref{f-idspec}) of HS\,2237+8154 clearly reveals the composite
nature of the object, with prominent TiO absorption bands in the red
part of the spectrum, hallmarks of a late type star, and broad Balmer
absorption lines in the blue part of the spectrum, characteristic of
the high-gravity atmosphere of a white dwarf. Narrow emission of \Ha\
and \Hb\ is detected, which is most likely associated with
chromospheric activity on the late-type companion
\citep[e.g.][]{bleachetal00-1}.

Additional time-resolved spectroscopy of HS\,2237+8154 was obtained in
August 2002 with the Intermediate Dispersion Spectrograph (IDS) on the
Isaac Newton Telescope, totalling 30 spectra spread out over one week
(Table\,\ref{t-obslog}). The IDS was equipped with the R632V grating
and the $2048 \times 4100$ pixel EEV10a detector. Using a slit width
of 1.5\arcsec the setup provided an unvignetted wavelength range of
$\sim4400\,\mbox{\AA}-6800\,\mbox{\AA}$ and a spectral resolution of
$\sim2.3$\,\AA. Copper-argon wavelength calibrations (arcs) were
obtained at the beginning and end of each observation block of
HS\,2237+8154.   The IDS data were
bias-subtracted and flat-fielded in a standard manner using the
\texttt{Figaro} package within the Starlink software
collection. Optimum extraction \citep{horne86-1} and sky line
subtraction was carried out using Tom Marsh's (\citeyear{marsh89-1})
\texttt{Pamela} package. The dispersion relation was established
fitting the arc line positions with a fifth-order polynomial. The rms
of the wavelength solution was smaller than 0.1\,\AA\ for all
spectra. Possible drift of the wavelength-to-pixel scale was accounted
for by interpolation between the arcs.

\begin{figure}
\centerline{\includegraphics[width=6.8cm]{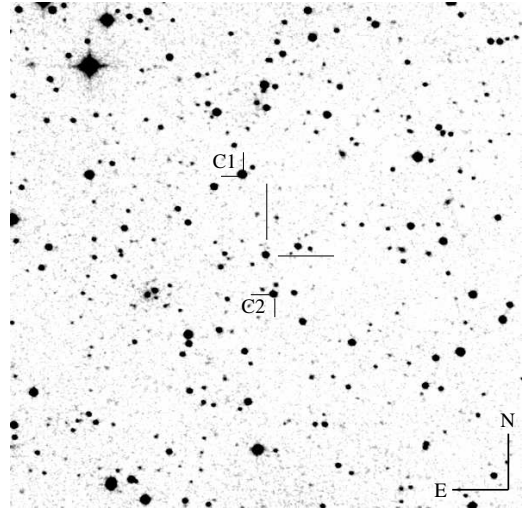}}
\caption[]{\label{f-fc} Finding chart ($10\arcmin\times10\arcmin$) for
HS\,2237+8154 obtained from the Digitized Sky Survey. The coordinates
of the star are $\alpha(2000)= 22^h37^m15.8^s$ and
$\delta(2000)=+82^{\circ}10\arcmin27.5\arcsec$. The comparison stars
used in the analysis of the Kryoneri data are marked 'C1'
(USNO--A2.0~1650--02587295 ) and 'C2' (USNO--A2.0~1650--02586211).}
\end{figure}

\begin{table}[t]
\caption[]{Log of the observations\label{t-obslog}.}
\begin{flushleft}
\begin{tabular}{rcccc}
\hline\noalign{\smallskip}
Date & UT Time &  Data & Exp.(s) & Num. Obs \\  

\hline\noalign{\smallskip}
\multicolumn{5}{l}{Spectroscopy} \\
1991 Jun 24 & 03:30-03:45 & $R\simeq6$\,\AA & 900 & 1 \\
2002 Aug 28 & 00:32-01:55 & $R\simeq1.6$\,\AA & 600 & 9 \\
2002 Aug 30 & 01:35-02:58 & $R\simeq1.6$\,\AA & 600 & 8 \\
2002 Sep 01 & 01:51-02:43 & $R\simeq1.6$\,\AA & 600 & 6 \\
2002 Sep 03 & 23:35-01:31 & $R\simeq1.6$\,\AA & 600 & 7 \\
\multicolumn{5}{l}{Photometry} \\
2002 Sep 19 & 18:31 - 23:40 &  $R$ & 10 & 961 \\ 
2002 Sep 20 & 17:42 - 22:03 &  $R$ & 10 & 960 \\ 
2002 Oct 08 & 17:16 - 22:17 &  $R$ & 10 & 982 \\ 
2002 Oct 17 & 17:38 - 21:52 &  $R$ & 30 & 415 \\ 
2002 Oct 18 & 17:00 - 23:18 &  $R$ & 45 & 417 \\ 
\noalign{\smallskip}\hline
\end{tabular}
\end{flushleft}
\end{table}

\subsection{Photometry}
We have obtained differential $R$-band photometry of HS\,2237+8154
during five nights in September/October 2002 at the 1.2\,m Kryoneri telescope
using a SI-502 $516\times516$ CCD camera (Table\,\ref{t-obslog}).
Bias and dark-current subtraction as well as flat-fielding of all CCD
images was carried out in a standard fashion within
\texttt{MIDAS}. Because of the relatively poor tracking none of the
standard photometry packages could be used to produce light curves
from the CCD data without excessive human intervention, and we
therefore set up the following reduction pipeline.
(1) We extracted detector coordinates and instrumental magnitudes for
all objects in each image using the \texttt{sextractor}
\citep{bertin+arnouts96-1} in the ``classic'' aperture photometry
mode. In order to account for variable seeing, this step is done as a
two-stage process: \texttt{sextractor} is run in a first pass using a
relatively large aperture diameter (10 pixel), the median seeing is
computed from the resulting object list, and then 
\texttt{sextractor} is run a second time, using an aperture diameter of
1.5~times the median seeing, which proved to give the best
signal-to-noise ratio. 
(2) One CCD image is designated to be the ``reference frame'', and the
object lists derived from all other images are matched against that of
the reference frame.  Considering that most of our CCD images of
HS\,2237+8154 have only three objects in common, and that no
significant distortion/scaling occurs between two different images, we
decided to implement the purpose-build matching algorithm
\texttt{smatch}. This program computes the distances between all
possible pairs of stars in a given image, and compares them with the
distances of all possible pairs of stars found in the reference
image. A ``match'' is found if the distances between two stars agree
within less than a given uncertainty $\epsilon$, with
$\epsilon<\mathrm{3\,pixel}$ in the case of the HS\,2237+8154
data. \texttt{smatch} proved to be robust for small numbers of stars
$(5-6)$ and provided a better match-rate than more complex codes such
as IRAF's \texttt{xyxymatch}.
(3) As a consequence of the small CCD detector in use at the Kryoneri
telescope (with a field of view of $2.5\arcmin\times2.5\arcmin$) and
the poor tracking, special attention has to be given to the choice of
the comparison stars, as some of them may move off the detector during
the observations. We therefore assign up to four different comparison
stars, and compute the differential magnitudes of the target star
relative to the brightest comparison visible on a given image.

$R$-band magnitudes of HS\,2237+8154 were derived relative to
USNO--A2.0~1650--02587295 ($R=14.5$) for the observations obtained in
September 2002, and relative to USNO--A2.0~1650--02586211 ($R=15.6$)
for the observations obtained in October 2002 (see
Fig\,\ref{f-fc}). During all five nights, HS\,2237+8154 was found at a
mean magnitude of $R=15.73\pm0.02$. However, it is important to keep
in mind that the \textit{systematic} error on this $R$ magnitude is of
the order $\sim0.2$\,mag due to the limited photometric accuracy of
the USNO--A2.0 catalogue (which lists HS\,2237+8154 with $B=16.1$ and
$R=15.9$). For completeness we note that $B=16.1$ has been derived from
the HQS observations of HS\,2237+8154.

\begin{figure}
\includegraphics[width=8.8cm]{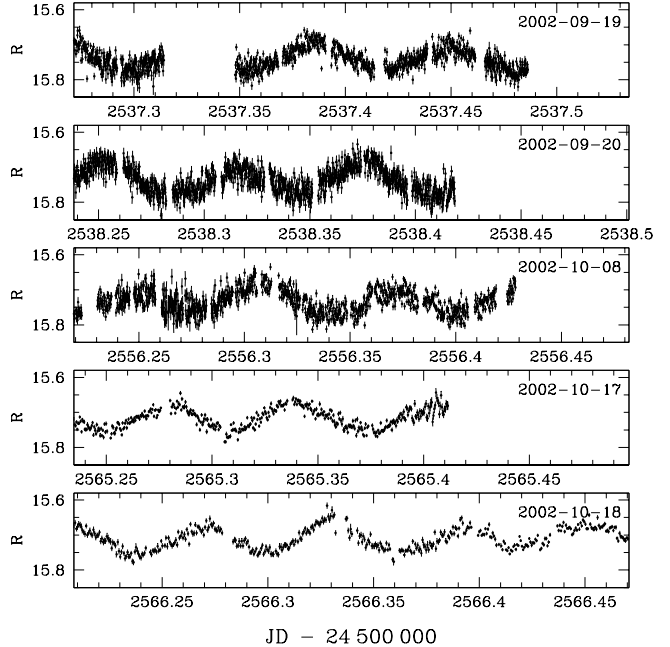}
\caption[]{\label{f-lc_all} Differential CCD $R$-band photometry
obtained at the Kryoneri 1.2\,m telescope. The $R$ magnitudes have
been computed relative to the comparison stars 'C1' (09/19 and 09/20)
and 'C2' (10/08 to 10/18). The data from the first three nights have a
lower signal-to-noise ratio because of the relatively short exposure
times (10\,s). }
\end{figure}

The light curves of HS\,2237+8154 (Fig.\,\ref{f-lc_all}) display a
quasi-sinusoidal variation of the $R$-band magnitude with a
peak-to-peak amplitude of $\sim0.1$\,mag and a period of $\sim90$\,min
during all five nights.

\section{The orbital period of HS\,2237+8154}

\subsection{\label{s-radvel}Radial velocities}
We have measured the radial velocity variation of the narrow \Ha\
emission line in the INT spectra using Gaussian fits. Care was taken
to avoid a contamination of the measurement by the broad M-dwarf
continuum peak located close to \Ha. The radial velocities were then
subjected to the analysis-of-variance algorithm
\citep[AOV][]{schwarzenberg-czerny89-1}, as implemented in the
\texttt{MIDAS} context \texttt{TSA}. The resulting periodogram
contains a number of peaks in the range $\sim2-10$\,\id, with the
strongest signal at 8.085\,\id\ (Fig.\,\ref{f-spect_power}). We have
created a faked set of radial velocities, computing a sine wave with a
frequency of 8.085\,\id and adopting the temporal sampling as defined
by the observed spectra. The errors of the observed radial velocity
measurements were used to randomly offset the faked radial velocities
from the value of the sine wave. The AOV analysis of this faked data
set is also shown in Fig.\,\ref{f-spect_power}. Whereas the
spectroscopy strongly suggests an orbital frequency (period) of
8.085\,\id\ (178\,min), it does not permit a secure identification.

\subsection{Photometry}
\begin{figure}[t]
\begin{minipage}{8.8cm}
\includegraphics[angle=-90,width=8.8cm]{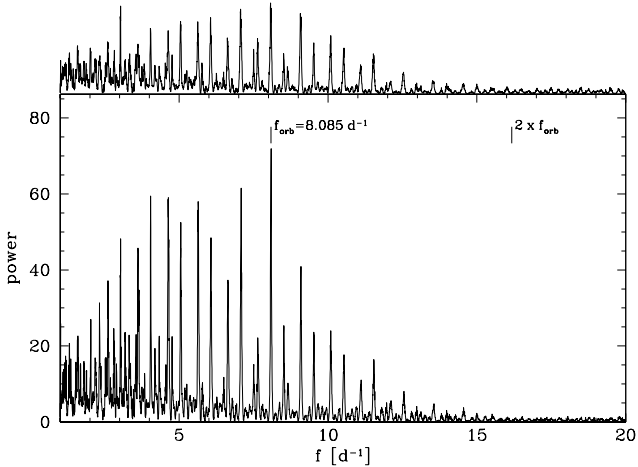}
\caption[]{\label{f-spect_power} Main panel: analysis-of-variance
periodogram of the radial velocities measured from the \Ha\ emission
line. Shown on top is the AOV periodogram of a faked set of radial
velocities (see text for details).}
\end{minipage}
\end{figure}

\begin{figure}[t]
\begin{minipage}{8.8cm}
\includegraphics[angle=-90,width=8.8cm]{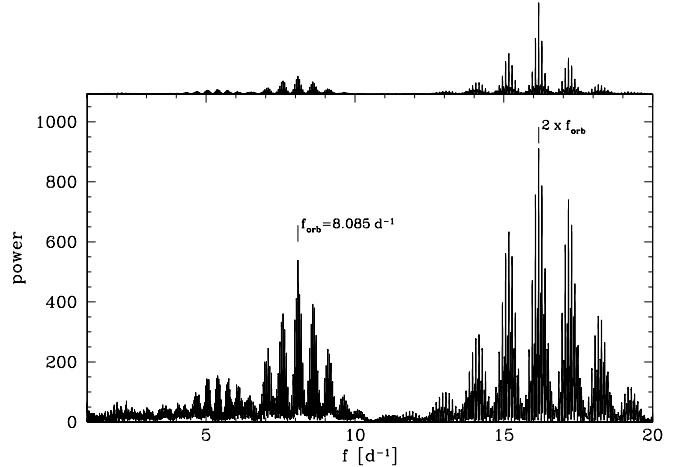}
\caption[]{\label{f-phot_power} Main panel: analysis-of-variance
periodogram computed from the five nights of differential photometry
shown in Fig.\,\ref{f-lc_all}. On top: the AOV periodogram computed
from a sine wave with $f=2\times\forb$ reproduces all structures
detected in the data.}
\end{minipage}
\end{figure}

The AOV periodogram computed from our photometric data of
HS\,2237+8154 is shown in Fig.\,\ref{f-phot_power}.  The strongest
peak is found at $f\simeq16$\,\id\ ($P\simeq90$\,min), which
corresponds to the period of the quasi-sinusoidal modulation apparent
in the light curves (Fig.\,\ref{f-lc_all}). A second structure in the
periodogram is centred at half that frequency, $\simeq8$\,\id\
($P\simeq180$\,min).  Considering that the $R$-band flux of
HS\,2237+8154 is almost entirely dominated by emission from the
secondary star, the observed quasi-sinusoidal variability is readily
explained by either ellipsoidal modulation, in which case
$\Porb\simeq180$\,min, or by reflection effect of the secondary
irradiated by the white dwarf, in which case $\Porb\simeq90$\,min. The
absence of a signal in the $\simeq16$\,\id\ range of the periodogram
obtained from the radial \Ha\ velocities (Fig.\,\ref{f-spect_power})
clearly rules out the second option. We therefore identify the
$\sim8$\,\id\ signal, which is detected consistently both in the
photometric variability as well as in the radial velocity variation,
as the orbital period of HS\,2237+8154.

Combining the photometric and spectroscopic observations, we define
the following ephemeris
\begin{equation}
\label{e-ephemeris}
\phi_0 = \mathrm{HJD}\,2452514.4112(1) +  0.12368(6)\times E
\end{equation}
were phase zero is the time of the blue-to-red crossing of the \Ha\
radial velocities. The most accurate measurement of the orbital
period, $\Porb=178.10\pm0.08$\,min, is derived from the photometry
which covers a longer time span compared to the spectroscopy.

\begin{figure}
\centerline{\includegraphics[width=8.8cm]{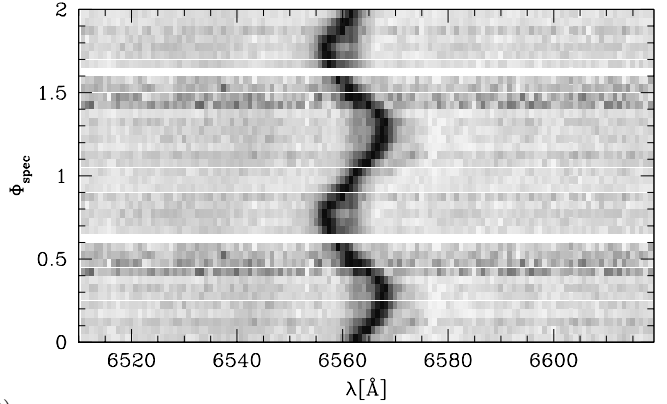}}
\caption[]{\label{f-trail} Trailed spectrogram of the \Ha\ emission
line in the IDS spectra of  HS\,2237+8154. The 30 individual spectra
have been normalised and averaged into 20 phase bins.}
\end{figure}

Figures\,\ref{f-phot_fold} and \ref{f-halpha_fold} show the Kryoneri
photometry and the INT radial velocities folded with this ephemeris.
The relative phasing of the photometry and spectroscopy is consistent
with our interpretation of the photometric variability as ellipsoidal
modulation: photometric minima occur during the zero-crossing of the
radial velocity curve (i.e. when the front/back side of the secondary
star is seen), photometric maxima occur during the radial velocity
maxima (i.e. when the secondary star is seen from the side). A trailed
spectrogram of the phase-folded IDS spectra around \Ha\ is shown in
Fig.\,\ref{f-trail}.

\section{The stellar components in HS\,2237+8154}
\subsection{Spectral fit}
We have fitted the identification spectrum of HS\,2237+8154
(Fig.\,\ref{f-idspec}) with a two-component model, consisting of
synthetic white dwarf spectra calculated with the code described by
\citet{gaensickeetal95-1}, and observed M-dwarf templates from
\citet{beuermannetal98-1}. We fixed the surface gravity of the white
dwarf models to $\log g=8$, corresponding to a canonical white dwarf
mass of $\sim0.6$\,\Msun, and allowed for temperatures in the range
$8000-25\,000$\,K. The spectral library of M-dwarfs extended from M0.5
to M9. The closest match of the observed spectrum was achieved for a
white dwarf with $\Teff=11\,500\pm1500$\,K and a dM3.5$\pm0.5$
companion (Fig.\,\ref{f-idspec}). The determination of the secondarys
spectral type is relatively robust, as the late-type star completely
dominates the emission of the system in the red part of the
spectrum. The flux from the white dwarf exceeds that of the companion
only for $\lambda\la5000$\,\AA, and consequently its temperature is
subject to a relatively large uncertainty. The scaling factors between
the model/template spectra and the observed spectrum of HS\,2237+8154
allow an estimate of its distance. From the white dwarf (assuming a
Hamada-Salpeter (\citeyear{hamada+salpeter61-1}) mass-radius
relation), we obtained $d=115\pm15$\,pc. Obviously, the distance would
be lower (higher) if the white dwarf were more (less) massive than
0.6\,\Msun. Assuming that the companion star fills its Roche volume
(see the discussion in Sect.\,\ref{s-discussion}), the scaling factor
of the M-dwarf template implies $d=95\pm15$\,pc. The agreement between
the two values is satisfying, and considering the overall
uncertainties involved in the spectral fit $d=105\pm25$\,pc appears to
be a conservative estimate, well within the reach of ground-based
astrometry programs \citep[e.g.][]{thorstensen03-1}.

\begin{figure*}
\begin{minipage}[t]{8.8cm}
\includegraphics[angle=-90,width=8.8cm]{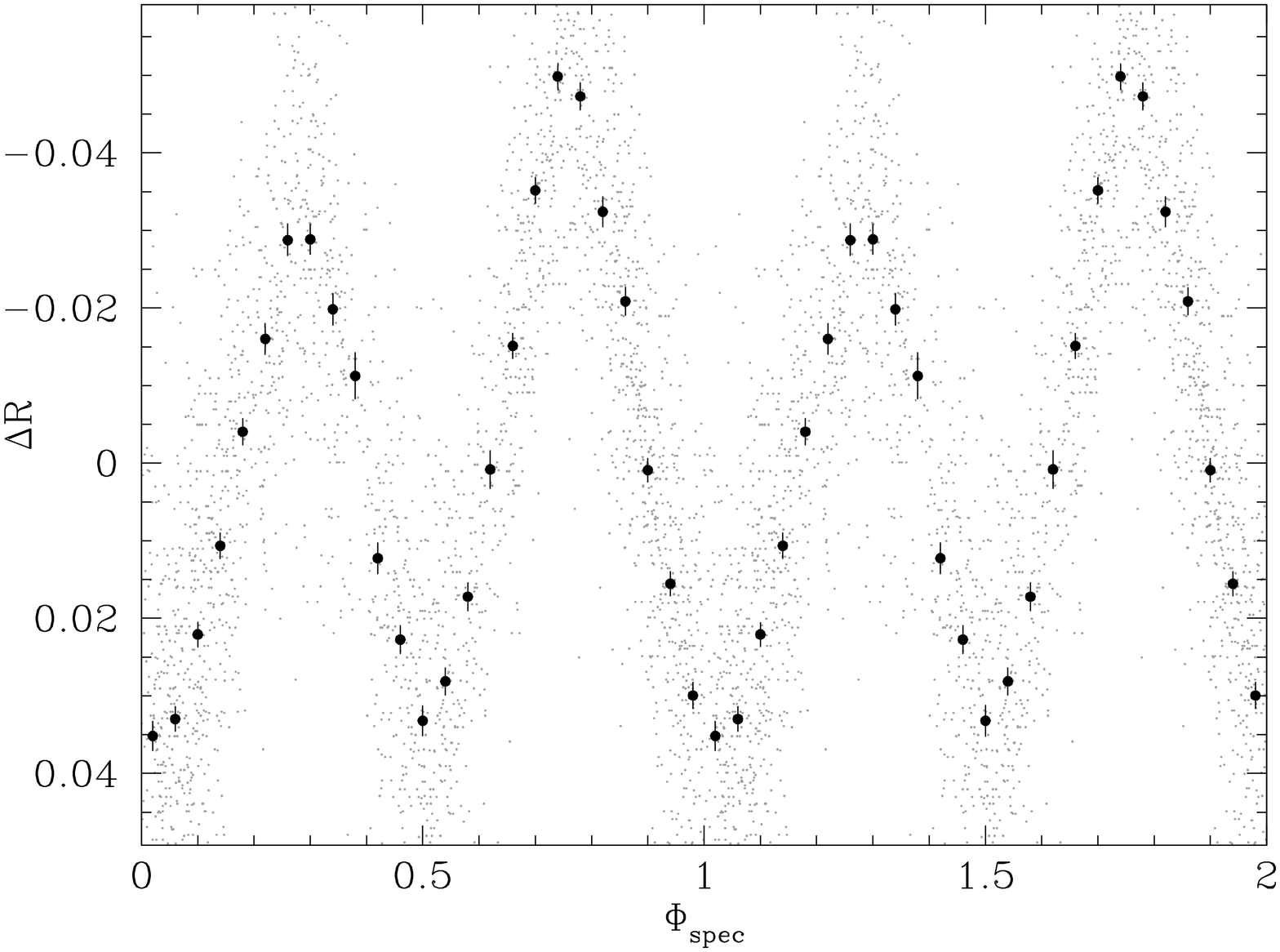}
\caption[]{\label{f-phot_fold} The Kryoneri photometry
average-subtracted and folded using the ephemeris
Eq.\,\ref{e-ephemeris} (gray points) and binned into 25 phase bins
(black points). Phase zero is defined as the blue-to-red crossing of
the \Ha\ radial velocity (see Fig.\,\ref{f-halpha_fold}).}
\end{minipage}
\hfill
\begin{minipage}[t]{8.8cm}
\includegraphics[angle=-90,width=8.8cm]{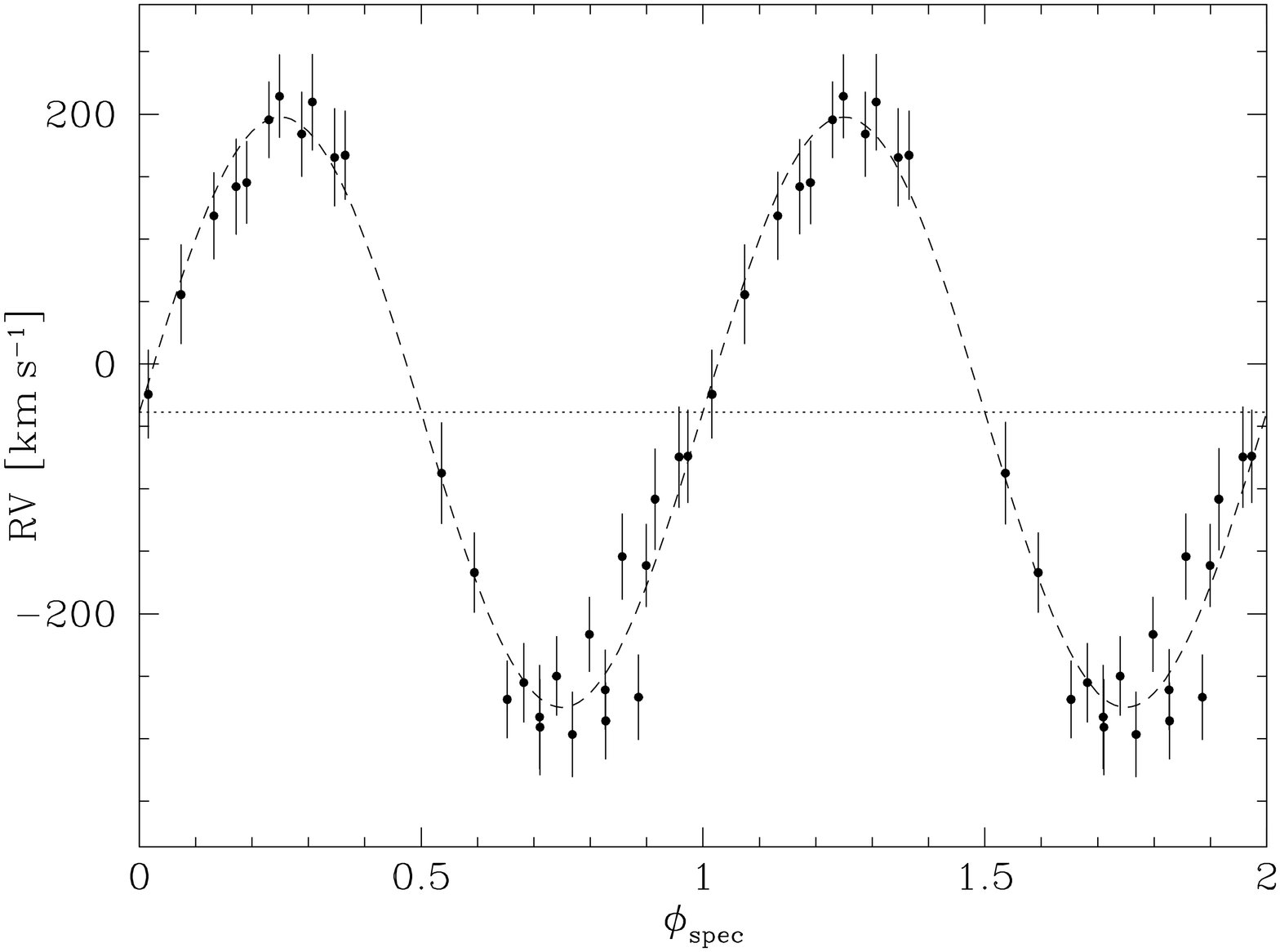}
\caption[]{\label{f-halpha_fold} Radial velocties measured from the
\Ha\ emission line, folded over the ephemeris given in
Eq.\,\ref{e-ephemeris}. Plotted as dashed line is a sine fit to the
radial velocity data which gives a semi-amplitude
$\Ksec=236\pm8$\,\kms\ and a systemic velocity
$\gamma=-40\pm6$\,\kms.}
\end{minipage}
\end{figure*}

\subsection{The radial velocity of the secondary}
Fitting a sine curve to the phase-folded radial velocities
(Fig.\,\ref{f-halpha_fold}) gives a semi-amplitude $\Ksec=236\pm8$\,\kms\ and
a systemic velocity $\gamma=-40\pm6$\,\kms. Adopting this value of
\Ksec\ implies a mass function 
\begin{equation}
\Mwd>f(\Mwd)=\frac{\Porb \Ksec}{2\pi
  G}=\frac{\Mwd\sin^3i}{(\Mwd+\Msec)^2}
=0.17\,\Msun
\end{equation}
Possible (\Mwd, \Msec) combinations that satisfy this mass function
are plotted for a variety of inclinations in
Fig.\,\ref{f-massfunc}. Inclinations $i\ga75^\circ$ are excluded from
the absence of eclipses (assuming that the secondary is nearly filling
its Roche lobe, see Sect.\,\ref{s-discussion}). The typical mass range
for field M-dwarfs with a spectral type dM3.5$\pm0.5$ is
$\Msec\simeq0.2-0.4$\,\Msun \citep{kirkpatricketal91-1,
baraffe+chabrier96-1, delfosseetal99-1}.

\subsection{Ellipsoidal modulation\label{s-ellipsoidal}}
The observed $R$-band light curves show an ellipsoidal modulation with
a peak-to-peak amplitude of $\simeq0.085$\,mag. No evidence for a
reflection effect is detected~--~not too big a suprise considering
that the white dwarf is rather cold. Intriguing is the fact that the observed
maxima in the $R$-band light curve are not equally bright. A similar
effect has been observed in the $R$-band light curve of the pre-CV
BPM\,71214 \citep{kawka+vennes03-1}, who suggested an inhomogenous
distribution of starspots as possible explanation. In fact, rapid
rotation due to tidal locking may result in enhanced activity on the
secondary stars of pre-CVs.

We have computed ellipsoidal modulation model light curves for a grid
of (\Mwd, \Msec, $i$) combinations using the analytical formulae of
\citet{binnendijk74-1}, adopting gravity and limb darkening
coefficients of $y=1.38$ and $x=0.80$ (the detailed choice of these
two coefficients does not significantly alter the results described
below). More specifically, we have searched (\Mwd, \Msec) combinations
that reproduce the observed amplitude of the ellipsoidal modulation
for a fixed value of $i$. In first order approximation this
implies a constant mass ratio $q=\Msec/\Mwd$. The resulting
constraints on (\Mwd, \Msec) are shown in Fig.\,\ref{f-massfunc} 
for $i=50^{\circ}$ and $i=70^{\circ}$.

Combining the constraints from observed amplitudes of the
ellipsoidal modulation and the $\Ha$ radial velocity variation
provides a lower limit on the Roche-lobe filling factor of the
secondary. Assuming the highest inclination and highest mass ratio
that are consistent with the mass function $f(\Msec)$
($i\simeq70^{\circ}$, $q=\Msec/\Mwd\simeq0.7$) implies
$\Rsec\ga0.75R_{L_1}$ (with $R_{L_1}$ the equivalent radius of a
Roche-lobe filling secondary). Decreasing $i$ and/or $q$ within the
constraints of $f(\Msec)$ requires \Rsec\ to be closer to $R_{L_1}$
($\Rsec\ga0.9R_{L_1}$ for $i\simeq60^{\circ}$, and $\Rsec\simeq
R_{L_1}$ for $i\simeq50^{\circ}$).

If the secondary star in HS\,2237+8154 has a mass that is typical for
its spectral type, and nearly fills its Roche volume, then the
combination of all constraints shown in Fig.\,\ref{f-massfunc} results
in conservative estimates of the binary inclination and of the
component masses: $i\simeq50^{\circ}-70^{\circ}$,
$\Msec=0.2-0.4$\,\Msun, and $\Mwd=0.47-0.67$\,\Msun.

\section{The evolutionary state of HS\,2237+8154\label{s-discussion}}
Based on its composite spectral appearence and orbital period
HS\,2237+8154 qualifies as a bona-fide pre-CV. However, the spectral
type of the secondary is conspicously close to that found in CV with
orbital periods around 3\,h \citep{beuermannetal98-1}.  Could
HS\,2237+8154 be a genuine CV caught in a state of low accretion? We
believe that this is rather unlikely for the following reasons. (1)
The system has been found near $B\simeq R\simeq16$ during six epochs
of observations.  (2) A number of quiescent dwarf novae are known to
reveal the photospheric emission of their stellar components during
quiescence. However their optical spectra are always significantly
contaminated by the accretion disc, in the form of strong Balmer lines
and some more or less noticeable contribution to the continuum
\citep[e.g.][]{szkodyetal00-3}. No  emission from a quiescent disc
is observed in HS\,2237+8154. (3) Many of the strongly magnetic
cataclysmic variables (polars) exhibit deep low states during which
their spectra are completely dominated by the stellar components
(e.g. AM\,Her \citealt{gaensickeetal95-1}; EF\,Eri
\citealt{beuermannetal00-1}). The non-detection of Zeeman-splitting in
the \Hb\ to \Hd\ absorption lines limits a hypothetical field of the
white dwarf in HS\,2237+8154 to $B\la5$\,MG , well below the lowest
field measured in polars.

\begin{figure}
\includegraphics[angle=-90,width=8.8cm]{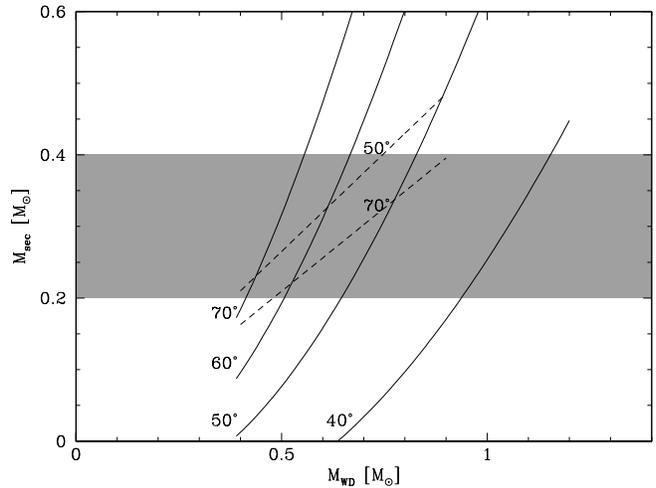}
\caption[]{\label{f-massfunc} Constraints on the binary inclination
$i$ and the stellar masses, \Msec\ and \Mwd. The gray shaded area is
the mass range typically found in isolated dM3.5$\pm$0.5 stars. The
solid lines indicate (\Mwd, \Msec, $i$) combinations consistent with
the measured mass function$f(\Msec)=0.17\,\Msun$. Plotted as dashed lines are
constraints derived from modelling the observed ellipsoidal
modulation.}
\end{figure}

Whereas the long-term behaviour and the spectral characteristics argue
against a genuine CV nature of HS\,2237+8154 the spectral type of its
secondary combined with the observed constraints from the observed
ellipsoidal modulation and radial velocity variation
(Fig.\,\ref{f-massfunc}) clearly show that the secondary star must be
nearly filling its Roche volume.

With a temperature of $\Teff=11\,500$\,K the white dwarf in
HS\,2237+8154 is unusually cold compared to the other known pre-CVs
and with an implied cooling age of $\sim400$ million years (for
$\Mwd\simeq0.6$\,\Msun) it is one of the oldest pre-CVs
\citep[see][]{schreiber+gaensicke03-1}. Taking the available results
at face value, we consider three possible evolutionary scenarios
for HS\,2237+8154. (1) It is a pre-CV close to
the onset of mass transfer. Ironically enough, it will be ``born''
close to the upper edge of the period gap. (2) It is a 
well-behaved CV~--~in the context of the disrupted magnetic braking model
of CV evolution~--~that turned off mass
transfer and entered the period gap. (3) It is a CV that is currently
hibernating in an extended low state, e.g. as a consequence of a
moderately recent nova eruption \cite{sharaetal86-1}.

It is interesting to note that two other pre-CVs have recently been
suggested to be hibernating novae as they contain nearly Roche-lobe
filling secondaries: EC\,13471--1258 \citep{odonoghueetal03-1} and
BPM\,71214 \citep{kawka+vennes03-1}. For these two systems, our option
(2) is not available, as they have $\Porb=217$\,min and
$\Porb=290$\,min, respectively.

\section{Conclusions}
We have identified a detached white/red dwarf binary with an orbital
period of $\Porb=178.10\pm0.08$\,min. Our spectroscopic and
photometric observations constrain the binary inclination to be 
 $i\simeq50^{\circ}-70^{\circ}$ and the stellar
masses to be $\Mwd=0.47-0.67$\,\Msun\ and
$\Msec=0.2-0.4$\,\Msun. Modelling the optical spectrum of
HS\,2237+8154 we find $\Teff=11\,500\pm1500$\,K for the white dwarf
and a spectral type dM3.5$\pm$0.5 for the secondary. The distance of
the system implied by the flux scaling factors of both stellar
components is $d=105\pm25$\,pc. 

The observations strongly suggest that the secondary star in
HS\,2237+8154 is nearly filling its Roche lobe, and we conclude that
the system is either a pre-CV just about to evolve into a
semi-detached configuration, a CV that evolved into a detached
configuration and entered the period gap, or a hibernating nova. All
three possible scenarios are extremely interesting
in the context of CV evolution, and we encourage future work to refine
the binary parameter of HS\,2237+8154.

\acknowledgements We thank Christos Papadimitriou and Dimitris Mislis
for carrying out part of the Kryoneri observations.  BTG was supported
by a PPARC Advanced Fellowship. SAB thanks PPARC for a studentship. We
are grateful to Darragh O'Donoghue and to Klaus Beuermann for
providing their M-dwarf templates. The HQS was supported by the
Deutsche Forschungsgemeinschaft through grants Re\,353/11 and
Re\,353/22. Tom Marsh is thanked for developing and sharing his
reduction and analysis packages Pamela and Molly.

\bibliographystyle{aa}

\end{document}